\documentclass[apsspec,twocolumn,showpacs,superscriptaddress,groupedaddress,nofootinbib,lengthcheck,preprintnumbers]{revtex4-2} 
\usepackage{graphicx}  
\usepackage{dcolumn}   
\usepackage{bm}        
\usepackage{amssymb}   
\usepackage{amsmath}
\usepackage{xspace}
\usepackage{epstopdf}
\usepackage{wrapfig}
\usepackage{color}
\usepackage{bbm}
\usepackage{float}
\usepackage{appendix}
\usepackage{subfigure}

\usepackage{times}
\usepackage{color}

\newcommand{\s}{$S$}
\newcommand{\g}{$X$}
\newcommand{\sbar}{$\bar{S}$}
\newcommand{\gbar}{$\bar{X}$}

\newcommand{\ssbar}{$S \bar{S}$}
\newcommand{\ggbar}{$X \bar{X}$}

\newcommand{\be}{\begin{equation}}
\newcommand{\ee}{\end{equation}}

\newcommand{\eehad}{$e^+\!e^-\! \!  \!\rightarrow \! \textit{hadrons}\,$}
\newcommand{\eeVgg}{$e^+e^-\! \rightarrow \! V  \! \rightarrow  \! X \bar{X}$}
\newcommand{\eeXhad}{$e^+e^-\! \rightarrow \! V  \! \rightarrow  \! X +\! \textit{hadrons}$}
\newcommand{\ecm}{ $E_{\rm cm}$}
\newcommand{\ep}{ $e^+e^-$}

\begin{document} 

\title{The muon g-2 and lattice QCD hadronic vacuum polarization \\ may point to new, long-lived neutral hadrons}

\author
{Glennys R. Farrar }
\affiliation{Center for Cosmology and Particle Physics,
Department of Physics,
New York University, NY, NY 10003, USA}


\begin{abstract}  

The experimental value the muon g-2 is $4.2 \sigma$ larger than the Standard Model prediction when the hadronic vacuum polarization contribution (HVP) is determined from the measured R-ratio, $\sigma(e^+e^- \! \rightarrow \! \mathit{hadrons})/\sigma(e^+e^- \! \rightarrow \!\mu^+ \mu^-)$; the HVP calculated in lattice QCD also significantly exceeds the measured R-ratio value.  We show here that both these discrepancies can be explained by an undetected contribution to $e^+e^- \! \rightarrow \! \mathit{hadrons}$ as could arise from production of previously unidentified neutral, long-lived hadrons.  We suggest two potential candidates for the new hadrons and propose several experimental tests.

\end{abstract}
\maketitle 

The magnetic moment of the electron has been of fundamental importance since Dirac's prediction that it should be twice the Bohr magneton.  The anomalous magnetic moment g-2 of the muon, the amount it differs  from the simple Dirac prediction, has become a fundamental test of particle physics because it  is sensitive to new particles or forces which are inaccessible to collider experiments, and it can be measured to extreme accuracy, a few parts in 10 billion.    A significant discrepancy, now $4.2 \sigma$, has emerged between the measured muon g-2~\cite{g-2PRL21} and the prediction from the Standard Model of particle physics~\cite{Aoyama+20}.  
The discrepancy, if persistent, would signal fundamental forces or particles not yet incorporated in the Standard Model or, as suggested here, yet-to-be discovered hadrons within the Standard Model.  

A crucial component of the Standard Model prediction of the muon g-2 is the determination of the Hadronic Vacuum Polarization (HVP) contribution, from experimental measurements of \eehad in the ``R-ratio" method.\footnote{The HVP contribution to the muon g-2 is given in the R-ratio method by
\be
\label{eq:amu}
a_\mu^{\rm HVP, LO} = \frac{\alpha^2 }{3 \pi^2} \int ds \frac{R(s) K(s)}{s}  ~,
\ee
where $s$ is the center-of-mass energy squared, $R(s) \equiv \sigma(e^+\!  e^- \! \rightarrow \! \textit{hadrons}) / \sigma_{\rm Born}(e^+\!  e^-\!  \rightarrow \! \mu^+ \! \mu^-)$  and $\sigma_{\rm Born}(e^+\!  e^-\!  \rightarrow \! \mu^+ \! \mu^-) =  4 \pi \alpha^2/ 3 s$; the function $K(s)$ is given in~\cite{Aoyama+20} and is approximately $m_\mu^2/3 s $.} 
Intriguingly, if the HVP value calculated in lattice QCD by the BMW collaboration~\cite{BMWHVP21} is used instead of the measured R-ratio value, the muon g-2 agrees with the SM prediction within the uncertainties.  Recently, Ref.~\cite{Alexandrou+ETMC22} -- combining the results of multiple lattice QCD groups -- sharpened the discrepancy with the experimental R-ratio to 4.2 $\sigma$.

Occam's razor suggests that rather than g-2 being anomalous \textit{and} the lattice QCD calculations being inconsistent with $e^+e^-$ data, coincidentally by the same amount, there may be a discrepancy between the measurement of $e^+ e^- \rightarrow \mathit{hadrons}$ and its true value.  We show here that this can be due to a class of undetected hadronic final states in $e^+e^-$ collisions.  The lattice QCD calculations sum over all states composed of quarks and gluons independently of their detectability, so would not be subject to the same deficit. Recast in this way, the muon g-2 and lattice HVP discrepancies are replaced by an ``R-ratio discrepancy" -- a difference between the R-ratio measured in $e^+e^-$ experiments and its true value when all hadronic final states are included.  

In this paper we show that certain types of hadronic final states in \ep\ collisions would have been systematically missed because of the stringent trigger, event-selection and analysis requirements imposed to suppress the severe backgrounds from beam-gas interactions and Bhabha scattering.  For instance, traditional triggers required at least one charged particle from the interaction region, or an energy deposit pattern in the detector consistent with photons from one or more $\pi^0$'s from the interaction point.  Such experiments exclude events with, for example, pair production of neutral long-lived hadrons, $X\bar{X}$, which carry a large fraction of the energy.  
We further identify a process which naturally fills the gap between the R-ratio prediction and the value derived from the observed muon g-2 and lattice HVP. 
In the following we 1) demonstrate that an R-ratio discrepancy due to missed hadronic states can naturally be at the root of the apparent g-2 and lattice anomalies, 2) suggest possible candidates for the required new particles and 3) propose experiments to test this scenario.
 
The R-ratio method applied to current $e^+ e^- \rightarrow \mathit{hadrons}$ data gives an HVP contribution $a_\mu \equiv \rm{(g-2)_{\rm had}}\times 10^{10} = 693.1 \pm 4.0$~\cite{Aoyama+20}.   This is $25.1\pm5.9$ less than the value required for the Standard Model prediction to agree perfectly with the measured g-2, and $13.7 \pm 5.5$ less than the BMW lattice QCD calculation.  Let us call this difference the HVP deficit, $\delta a_\mu \approx 20$. 
About 75\% of the HVP calculated from the R-ratio is due to the $\rho$ meson and 6\% to the $\phi$ meson, so the HVP deficit, $\delta a_\mu / a_\mu \approx 3\%$, is about half the size of the $\phi$ meson contribution.  A natural solution to the HVP deficit would thus be production of another vector meson, here designated $V$, which decays almost exclusively to undetected final states $X \bar{X}$.  We give below several examples which fill the HVP gap.  Then we review the event selection criteria of \eehad\ experiments, establishing that $X \bar{X}$ production would have been missed.   The $V\rightarrow X \bar{X}$ scenario is natural and its consistency with experiments can be verified without a dedicated detector simulation, but associated production, \eeXhad, can also explain the R-ratio deficit and also evades event selection requirements of current experiments, as long as a sufficiently small fraction of the CM energy goes into ordinary particles; detailed study of these more complex final states is left to future work. 

In order to explain \textit{both} the muon g-2  \textit{and} the lattice HVP anomalies, the missed events must contain particles present in the lattice gauge calculation, i.e., they must be made of $u,d,s$ or $c$ quarks, antiquarks and gluons.  
The lattice HVP calculation can correctly incorporate the contribution of hadronic states without those states having yet been explicitly identified via lattice QCD calculations.  This is because the lattice HVP calculation sums over intermediate states and is therefore inherently more robust than lattice predictions for individual states, much as perturbative QCD can accurately describe Deep Inelastic Scattering while individual production cross-sections are still not calculable.  Thus an HVP excess would naturally be the first indication of undiscovered neutral long-lived states in the hadron spectrum.\footnote{Neutral, because charged hadrons could not have escaped detection, and long-lived in order not to have been discovered by displaced vertex searches; see below. }  Future, more accurate lattice HVP calculations will identify the mass range predominantly responsible for the excess with respect to the observed R-ratio contribution~\cite{Colangelo+HVPwindow22,Davies+halfWindow22,Ce+HVPwin22} and hence narrow the mass region to be searched -- both experimentally and on the lattice -- for the new states.  We can anticipate that the new states will have mass $\lesssim $ few GeV, since non-perturbative QCD effects are most prominent at low energy and moreover the contribution of a state to the HVP decreases with increasing mass. \\

\noindent  {\bf \textit{Production of $X \bar{X}$}}

One of the ways that \ggbar\ pairs can be produced in \ep\ collisions is through a vector meson. Such contributions are well-described as Breit-Wigner resonances characterized by their mass, width and branching fractions into the initial and final states.  We shall use the $\rho^0$ and $\phi$ and their charmonium and bottomonium analogs, $\psi(3770)$ and $\Upsilon(4S)$, for guidance in modeling  \ggbar\ production via \eeVgg.  All these are $J^{PC} =1^{--}$ resonances (as  required to match the virtual photon quantum numbers) which decay predominantly to two spin-0 particles.  We also explore the possibility that \g\ has spin-1/2. 

Following the analysis of the SND collaboration at VEPP for $e^+ e^- \rightarrow \phi \rightarrow K_L K_S$~\cite{achasov+phi01}, we write
\be
\label{eq:sigV}
\sigma_{e^+ e^- \!\rightarrow \!V \!\rightarrow \!X \bar{X}}(s) \!=\! 12 \pi \!\left(\frac{q_X(s)}{q_X(m_V^2)}\right)^{2L+1} \! \!\frac{\Gamma_X\, \Gamma_{e} \, (m_V/\sqrt{s})^5 }{ [(m_V^2 - s)^2 + \Gamma_V^2 \, s]}~,
\ee
where $q_X(s) \! \equiv \! \sqrt{s/4 - m_X^2}$ is the magnitude of the center-of-mass 3-momenta of \g\ and \gbar\ at $s=E_{\rm CM}^2$ and $L$ is the orbital angular momentum of the final \ggbar. $L=1$ is required to conserve angular momentum when \g\ is a scalar or pseudoscalar, while $L=0$ for $X$ having spin-1/2.   
The parameters of the Breit-Wigner are $m_V$, $m_X$, the total width $\Gamma_V$, and the partial widths $\Gamma_X$ and $\Gamma_{e} $,  for $V$ decay to \ggbar\ and \ep.  The $V$ cannot have a significant probability of decaying to visible hadrons or we would already know of its existence, so to good approximation $\Gamma_X = \Gamma_V$ since $\Gamma_e << \Gamma_V$.  Allowing for multiple resonances is a straightforward extension.  

\begin{figure}[t]
     \centering
     \includegraphics[width=0.48\textwidth]{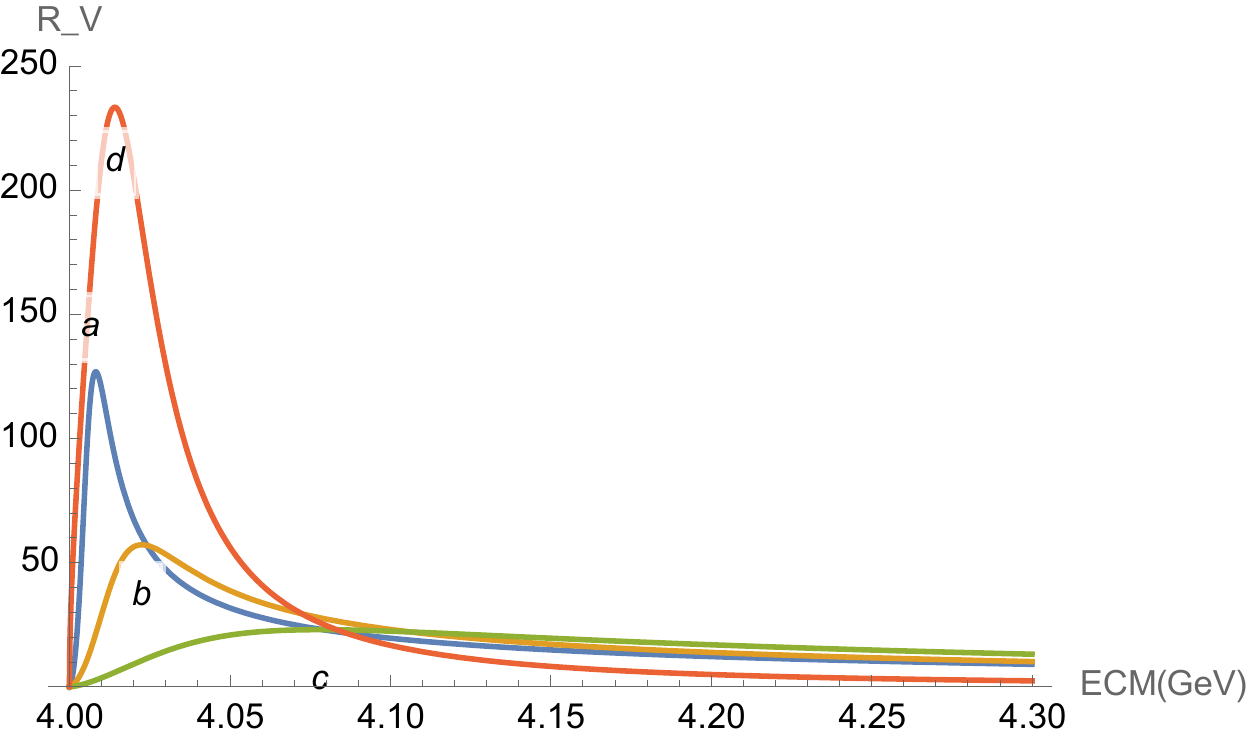}
        \caption{The undetected hadronic contribution to the R-ratio from $V$ decay to \ggbar, for four illustrative models taking $m_X = 2$ GeV for definiteness.  Details of the models are given in Table~\ref{tab:models} and the scaling with mass and other parameters is discussed in the text.  Cases a-c are for spin-0 and d (shown in red) is for spin-1/2.    \label{fig:RV}}
\end{figure}

Figure~\ref{fig:RV} shows $R_V(s)$, the increment in the R-ratio due to \ggbar\ production, for four choices of mass and width of the resonance $V$ 
selected for making certain points in the discussion of experimental tests below. For each example, $\delta a_\mu = 20.2$ and the combined fit to the muon g-2 and the BMW HVP has $\chi_{\rm dof}^2 = 0.8$; Table~\ref{tab:models} lists the details of each case. 
The widths adopted are representative of values for the vector mesons $\{\rho^0,\, \phi,\,\psi(3770), \,\Upsilon(4S)\}$, for which $\Gamma_V = \{147, 4, 27, 20 \}$ MeV,  $\Gamma_e = \{7, 1.3, 0.3, 0.3 \}$ keV,  and $\int \!\sigma_V dE_{cm}  = \{278, 31, 0.75, 0.25 \}$ nb-GeV.  This shows that a vector meson with properties like those we know, decaying predominantly to a previously unobserved long-lived neutral, naturally gives rise to the observed increment in the R-ratio.   There could also be multiple $V_i,X_i$ systems making individually smaller contributions.

These examples take $m_X = 2$ GeV but they can be rescaled to other $m_X$ values: for a fixed $\Gamma_{e}$, $\delta a_\mu$ scales approximately as $(m_X\, \rm{or}\, m_V)^{-2.55}$. The contribution of \eeVgg\ to $a_\mu$ depends weakly on the separation between resonance mass and threshold, $m_V - 2 m_X$, unless this becomes larger than the width of $V$, and it increases slower than linearly with increasing $\Gamma_V$.  The distinctly higher peak in $R_V$ seen in Fig.~\ref{fig:RV} for the spin-1/2 case is because the Breit-Wigner damps more rapidly at high energy for $L=0$ than for $L=1$, but inclusion of form-factor effects for the $L=1$ case would reduce the difference.  As the last two columns of Table~\ref{tab:models} show, the integrated contribution is similar in all cases.   

\begin{table}[t]
\label{tab:models}
\footnotesize
\begin{tabular}{ | c | c | c | c | c | c | c | c |}
\hline
case & L & $m_V$  & $m_X$  & $\Gamma_V$ & $\Gamma_e$ & $\int R_V dE_{cm} $  & $\int \! \sigma_V dE_{cm} $  \\
\hline
\hline

a & 1 & 4.005 GeV & 2 GeV & 10 MeV & 5 keV & 12.1 GeV & 56.8 nb-GeV\\
b & 1 & 4.01 GeV & 2 GeV & 30 MeV & 5.1 keV & 12.4 GeV & 57.1 nb-GeV\\
c & 1 & 4.003 GeV & 2 GeV & 100 MeV & 0.4 keV & 13.2 GeV & 57.7 nb-GeV\\
d & 0 & 4.01 GeV & 2 GeV & 30 MeV & 38 keV & 10.3 GeV & 55.4 nb-GeV\\
\hline\end{tabular}
\caption{{ \footnotesize   The parameters of the 4 illustrative models shown in Fig.~\ref{fig:RV}.  The last column shows the integrated cross section $\int \sigma_{e^+e^-\!\rightarrow V} dE_{cm} $.}} \label{tab:models}
\vspace{-0.15in}
\end{table}

\vspace{0.08in}
\noindent {\bf \textit{ Experiments}}

Here we review the most important $e^+e^-$ experiments, historically and for the R-ratio determination, to ascertain whether \ggbar\ final states would be included.   We do not include experiments measuring a specific final state such as $K^+K^-$, which are not relevant for this discussion. Reference~\cite{Aoyama+20} gives an overview of all of the experiments, as well as the methodology used in the R-ratio determination of the HVP. \\
$\bullet$  {\bf Mark I} at SPEAR,  the experiment that discovered the $\psi$,  reported measurements of the total cross section for hadron production in \ep\ collisions between 2.6 and 7.8 GeV~\cite{MarkI82}.  They only recorded events with at least two charged particles, and further required reconstructed tracks to intersect in the collision region.   
\\
$\bullet$  {\bf The Crystal Ball} collaboration measured the total hadronic and D* cross sections in the range $3.87 \,{\rm GeV} < E_{\rm CM} < 4.5$ GeV~\cite{OsterheldCrystalBall86}.  The heart of the Crystal Ball was a hermetic (93\% of $4 \pi$) array of NaI crystals in which photons and electrons deposit virtually all of their energy.  An inner tracking system tagged charged particles.  In order to qualify as a hadronic candidate, an event without charged tracks had to i) have a total energy deposit of at least 35\% of \ecm\ and ii) pass a \textit{``maximum asymmetry cut"} of 0.85, such that along no axis is the difference in energy between hemispheres greater than 15\% of the total observed energy.  
These event selection criteria mean that \eeVgg\ events would generally not be detected.  Condition i) is not met except for possible instances of \gbar\ annihilation, because the total energy deposit is limited by the kinetic energy of the \ggbar: $m_V - 2 m_X \lesssim \mathcal{O}(100 \, {\rm MeV})  << 0.35 \, E_{\rm CM} \approx 1.3 $ GeV.  And in case \gbar\ carries baryon number -1 or lower and annihilates in the detector, condition ii) would generally cause the event to be rejected on grounds of too-large energy asymmetry:  in the direction of the \gbar\ annihilation at least $ m_X + m_N $ less the mass of the final state products would be released, while in the opposite hemisphere the maximum KE is $m_V/2 - m_X,~\mathcal{O}(10's \, {\rm MeV})$.\\
$\bullet$  {\bf BES}, the Beijing Spectrometer at the Beijing Electron Positron Collider (BEPC), is the most powerful instrument now measuring \ep\ collisions below a CM energy of around 5 GeV.   BEPC is a symmetric \ep\ collider producing a very high luminosity.  The BESII collaboration took data at 85 CM energies between 2-5 GeV (34 energies between 3700 and 4200 MeV with roughly 10 MeV spacing, closer near known structures)~\cite{BES02};  results from the BESIII energy scan have not yet been published.  Event selection criteria for BESII were mainly as described in Ref.~\cite{BES00}.  They require an event to have at least two charged particles and a total deposited energy in the EM calorimeter of at least 0.28 $E_{\rm beam}$, so they would exclude a  $V\rightarrow X \bar{X}$ process.  An additional requirement is that events with just two charged particles must have at least two additional neutral clusters of more than 100 MeV in the EM calorimeter, not too co-linear with the charged tracks.  (The neutral clusters could be produced by a photon or neutral particle such as $K_L$ or neutron). \\

The experiments described above used the ``scan method" of stepping through the energy range with steps small compared to structure in the cross section.  Another approach is the Initial State Radiation (ISR) method, in which the CM collision energy $\sqrt{s}$ is typically fixed to a resonance such as an  $\Upsilon$, but the CM energy of the \ep\ annihilation process takes a continuous value over a large range due to emission of a hard ISR photon.   The effective \ecm\ of the \ep\ annihilation in the ISR method is completely determined by the accelerator collision energy, $\sqrt{s}$,  and the ISR photon energy in the CM,  $E_\gamma^*$: 
$
E_{\rm cm}^2 = s - 2 E^*_{\gamma} \sqrt{s}.
$
Thus the ISR method has the advantage of not missing a narrow resonance between steps in the scan, although limitations in measuring $E_\gamma^*$ limit the resolution in $E_{\rm cm}$.   The price paid with the ISR method is less well-determined collision kinematics and smaller cross section.    In principle -- with sufficiently good ISR photon energy resolution -- the ISR method provides an ideal approach to performing a completely inclusive search which can discover an otherwise invisible contribution to $\sigma(e^+ e^- \rightarrow $ \textit{everything}).  However up to now the only use of the ISR method, in the context of measuring $R(s)$, has been for exclusive final states.\\

$\bullet$  {\bf BaBar}, at the SLAC asymmetric B-factory, was sensitive to $V\rightarrow X\bar{X}$ via direct production for $m_V$ around 10 GeV, and through ISR for lower masses~\cite{LeesBABAR13}.    An in-depth study of the inclusive ISR capabilities with BaBar is given in an unpublished thesis~\cite{bergerISR06}.  
Although  BaBar has not yet reported an inclusive total \ep\ cross section, they have used the ISR technique to place limits on the kinetic mixing parameter between the QED photon and a possible massive ``dark photon"~\cite{BABARdark17}.   That analysis fits the observed spectrum of ISR photons to a zero-width resonance plus a smooth background of unspecified shape.  Their event selection aimed to identify events with exactly one ISR photon and no other activity, hence demanded that the ISR photon not be accompanied by any charged particles or any neutral particle energy deposit.  
The\ BaBar dark photon search can be directly applied to constrain the \eeVgg\ scenario if the $V$ is narrow and the \g\ and \gbar\ do not interact.  However neither of these conditions is generic to the missing hadrons scenario.  As a first step, removing the requirement that the \g\ or \gbar\ not interact in the detector could test the examples given in Fig.~\ref{fig:RV} when the width of $V$ is small compared to the detector resolution.  \\

To summarize, events in which the energy is dominantly carried by new neutral hadrons would not have been identified as an $e^+e^-\rightarrow \textit{hadrons}$ event by experiments to date, unless the $X  \bar{X}$ were accompanied by sufficiently energetic charged particles or deposit a significant fraction of their own energy in the detector.  Other types of events can also escape detection, but this must be established on a case-by-case basis. \\

\noindent {\bf \textit{ How could an \g\ particle not be already known? }}

The discussion above and the \eeVgg\ model demonstrate that certain contributions to the R-ratio, sufficient to remove the discrepancy with g-2 and lattice QCD, would not have been detected.  Events in which most of the energy is carried by long-lived neutral hadrons do not generally satisfy event selection requirements.  However since the hadron spectrum is very well studied, it seems surprising that there could be unknown long-lived hadrons.  The minimal conditions for a state \g\ to have been missed are that it is neutral, not part of a multiplet with charged members which have been detected, and long-enough lived not to have been noticed through its decays.  A lifetime $\tau \gtrsim 10^{-7}$s is probably sufficient for this, but a comprehensive examination of experiments is warranted. 

If $m_X$ is less than a few GeV, and the product of the $X$ production cross section in hadron collisions and its scattering cross section on nuclei is small compared to that product for neutrons, it would escape notice in high energy hadron scattering and collider experiments: a background of vastly more neutrons is a potent camouflage!  Indeed, the neutral particle search of Ref.~\cite{gustafson} explicitly excluded masses below 2 GeV.  

As a hadron, the \g\ is presumably mildly or moderately strongly-interacting, so it loses energy in matter and would not be penetrating enough to be seen in searches such as FASER~\cite{FASER21} or MilliQan~\cite{milliQan21}.  Indeed, discovering a mild-to-moderately interacting long-lived neutral particle in the GeV-range requires a dedicated effort.  A strategy and detector to find a subdominant population of long-interaction-length-neutral particles is outlined in Sec.~VII.E of Ref.~\cite{fS22}.\\

\noindent {\bf \textit{ What could the \g\ particle or particles be? }}

The biggest impediment to the existence of a suitable candidate for \g\ is the requirement that it be long-enough-lived.  A long lifetime requires some mechanism for stabilization.

Phenomena such as instantons, sphalerons and Skyrmions in particle physics, and topological order and vortices in condensed matter systems, provide motivation to consider whether QCD may support a not-yet-identified topologically-stabilized state.  If such a state exists, it might be resemble an exotic, conceivably fermionic glueball-$\eta'$ hybrid, suggesting a mass in the 1-2 GeV range.  If it were actually spin-1/2, the $X$ would have to be pair produced along with an $\bar{X}$, as already envisaged in the \eeVgg\ model.   A concrete model would enable evaluation of whether production of such a particle could explain the HVP deficit.

A less speculative mechanism for a hadron to have a long lifetime is if its constituent quarks carry a quantum number that is conserved. A candidate of this type appears in the  stable sexaquark scenario~\cite{fS22}, where \g\ would be the scalar, flavor-singlet di-baryon \s, composed of $u_\uparrow u_\downarrow d_\uparrow d_\downarrow s_\uparrow s_\downarrow $.  Here, the conservation of baryon number, and of strangeness by the strong interactions, leads to a very long lifetime if the \s\ is light enough. For consistency with the stability of the deuteron but otherwise being minimally restrictive, the mass range of the \s\ is 1850 MeV $< m_S \lesssim  2230$ MeV.\footnote{For $ m_S \lesssim  2050$ MeV, the lifetime exceeds the age of the Universe~\cite{fzNuc03}.  See Ref.~\cite{fS22} for more information about such a state and a discussion of why it would have escaped detection, as well as suggestions for experimental approaches to discovering it.}
The \eeVgg\ mechanism can potentially be realized with $X$ being the sexaquark.  The exchange of the flavor-singlet combination of $\omega$ and $\phi$ vector mesons produces an attractive Yukawa interaction between \s\ and \sbar.  Explicit numerical solution of the Schroedinger equation shows that  $1^{--}$ bound states are formed for reasonable coupling, with width and mass offsets compatible with the examples of Table I.\footnote{X. Xu, private communication.}  

However $e^+ e^-\!\rightarrow \!V\! \rightarrow S \bar{S}$ may not be the dominant production channel.  According to the Vector Meson Dominance model, $\Gamma_{e}$ is proportional to the square of the mean charge at the origin.  Although the \s\  would have a non-zero charge radius since the strange quark is significantly heavier than the light quarks, angular momentum conservation requires the \ssbar\ in the $V$ to have $L=1$, suppressing the charge at the origin.  Thus associated production, e.g., $e^+e^- \rightarrow  \bar{S} \Lambda \Lambda $ or $S \bar{\Lambda} \bar{\Lambda}$, could be more important than $e^+e^-\! \rightarrow \! V  \! \rightarrow  \! S \bar{S}$, especially in view of the large production cross section $\sigma(e^+ e^- \rightarrow \Lambda \bar{\Lambda})$ near threshold~\cite{ZhouLamLam22}.  Associated production might have evaded detection because the BESII event selection requires an event with two charged tracks to have accompanying neutral clusters, and the Crystal Ball anti-asymmetry cut would exclude such events since the rest mass of the \s\ would not be detected.  
Dedicated study with an event generator is needed to delimit this possibility.  

An intriguing anomaly reported in~\cite{bergerISR06} is a potential hint of \sbar\ production: BaBar sees an excess of high energy anti-protons relative to simulation, with no such discrepancy for protons, as shown in Fig. 9.34 of~\cite{bergerISR06}.  This could be a signature of anti-sexaquark annihilation producing $\bar{\Lambda}$, whose decay produces the observed $\bar{p}$ excess.  (The generic suppression of reactions involving \s\ interconversion with baryons at low momentum transfer~\cite{fS22} is mitigated by BaBar's asymmetric collider configuration causing possible \sbar's produced at rest in the CM to have $\beta \approx 0.5$ relative to the detector.)  \\

\noindent  {\bf \textit{Predictions and Tests}}

Several experimental approaches can shed light on the possibility of an unseen contribution to \ep\ final states:  
\vspace{0.08in}\\
$\bullet$  {\bf Measurement of the $e^+ e^- $ total cross section using ISR} with existing BaBar data, or potentially at Belle II, is an ideal approach if event selection can be clean enough and background processes adequately modeled.  The condition of no neutral energy deposit should be removed to avoid suppressing the signal of interacting neutrals, since hadrons generically will interact.  Good resolution on $E_{cm}$ of the $e^+e^-$annihilation reaction would be valuable to trace the structure of the cross section and expose differences in shape relative to the measured (``visible") $R(s)$, enabling the mass to be constrained.  A corollary of the missed-events scenario is that excess in the ISR inclusive measurement should be associated with a corresponding population of events that would not have passed selection cuts for the standard R-ratio measurements. 
Seeing a sufficient excess in inclusive ISR over expectations from the visible R-ratio measurements would validate the proposed mechanism of an ``invisible" contribution to $e+e-$ final states as the solution to the g-2 anomaly, but demonstrating the existence of such an invisible component is not alone sufficient to discriminate between the production of unseen hadronic states or Beyond the Standard Model particles.  In the latter case, the lattice HVP anomaly would not be explained.  \vspace{0.01in}

\noindent $\bullet$  {\bf Precision measurement of $e^+ e^- \rightarrow \mu^+ \mu^-$}, using the ratio $R_{\mu \mu} \equiv \sigma_{e^+ e^- \rightarrow \mu^+ \mu^-}/ (4 \pi \alpha^2/ 3 s)$ to probe the vacuum polarization as proposed  in Ref.~\cite{Dong+CZY20} and done in Ref.~\cite{BESmumu20}.  The present statistical precision on $R_{\mu \mu}$ is somewhat better than 1\% in the scan data and ~0.2\% in high luminosity data, but the systematic uncertainty is 2.9\%; the latter is expected to improve significantly, possibly to better than 1\%.  
\begin{figure}[t]
     \centering
     \includegraphics[width=0.48\textwidth, trim=0.7in 1.3in 0.3in 0.8in,clip]{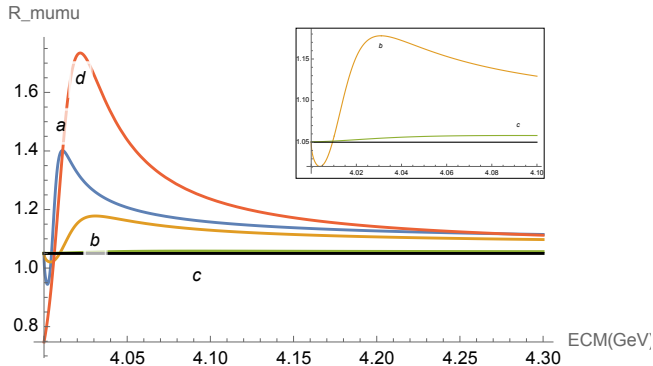}
        \caption{The predicted ratio $R_{\mu \mu}$ for the four illustrative cases of Fig.~\ref{fig:RV}, taking $R_{\rm had}$ without $R_V$ to give $R_{\mu \mu} = 1.05$.  Cases requiring large $\Gamma_{e}$ to fit the HVP should be testable via $R_{\mu \mu}$.  The inset is a blow-up for cases b and c.         \label{fig:Rmumu}}
        \vspace{0.15in}
\end{figure}
To illustrate the discriminating power of a precision measurement of $R_{\mu \mu}(s)$, Fig.~\ref{fig:Rmumu} shows the impact of \eeVgg\  on $R_{\mu \mu}$ in the four models introduced previously  to explore solution space   (Fig.~\ref{fig:RV} and Table~\ref{tab:models}), approximating the visible $R(s)$ in this region as giving a constant $R_{\mu \mu} = 1.05$ in the absence of a $V$:\footnote{I thank C. Yuan for this suggestion, as well as valuable discussions about BESIII experimental capabilities.} 
\be
\label{eq:Rmumu}
R_{\mu \mu} \!=\! \rm{Abs}\! \left[ 1.025 + \frac{3 s}{\alpha} \! \left(\frac{q_S(s)}{q_S(m_V^2)}\right)^L \! \frac{k_{ee} \, \Gamma_{e}/m_V}{(s - m_V^2 + i m_V \Gamma_V)} \right]^2 .
\ee
Figure~\ref{fig:Rmumu} shows that some resonance properties for \eeVgg\ which solve the HVP discrepancies should already be excludable by BESIII data -- particularly those requiring large $\Gamma_e$ like models a and d.  But model c underlines the fact that an invisible contribution to the R-ratio would be difficult to see in $R_{\mu \mu}$, if $\Gamma_e$ is small.  \vspace{0.08in}

$\bullet$  {\bf Search for \ggbar\ production} through interactions in the detector, analogous to the BESIII measurement of $n \bar{n}$ production in the 2-3.08 GeV range~\cite{BESneutron21}.  The \ggbar\ search would be more challenging because the $n \bar{n}$ study exploited the subset of events in which the $\bar{n}$ annihilates, producing a characteristic asymmetry in the energy deposit in opposite directions.  However annihilation cannot be relied on in either  of the long-lived $X$ scenarios discussed above:  a  topologically stabilized state could not annihilate with a nucleon because of baryon number conservation, and annihilation is much suppressed relative to scattering for an \sbar, at least at low energy~\cite{fS22}.  So the energy deposit on opposite sides of the detector would be on average symmetric, and limited by $(E_{\rm CM} - 2 m_X)/2$.  Running at the highest beam energy to maximize the energy the \ggbar\ can deposit, would  produce the most visible signal.\\

\noindent  {\bf \textit{Summary} }

The experimentally measured value of the anomalous magnetic moment of the muon~\cite{g-2PRL21} is 4.2$\sigma$ above the prediction of the Standard Model, if the Hadronic Vacuum Polarization contribution is inferred from precision measurements of cross sections for $e^+ e^- \rightarrow \mathit{hadrons}$ using the R-ratio method~\cite{Aoyama+20}. The latest lattice QCD calculations of the HVP~\cite{BMWHVP21,Alexandrou+ETMC22} also exceed the R-ratio value at a similar level.  
We have proposed here that both of these discrepancies may have a common resolution via an undetected contribution to \ep$\rightarrow$ \textit{hadrons}.   We showed that, given the event selection and trigger requirements of the \ep\ experiments to date, production of final states consisting only of new neutral particles \ggbar, or such neutrals with a sufficiently discreet accompaniment of ordinary hadrons, would have gone undetected.  
Hypothesizing the existence of such a hadron \g, we showed as a concrete ``proof-of-principle" that production of \ggbar\ pairs through a vector meson $V$ naturally reconciles the R-ratio, lattice QCD and muon g-2 results.

To provide a viable explanation for the muon g-2 and lattice HVP puzzles, the new state \g\ must be composed of $u,d,s,c$ quarks and gluons.  Furthermore it must be sufficiently long-lived not to have already been discovered.  We identified two potential candidates for the required long-lived neutral particle \g:  a conjectural topologically-stabilized hadron and the stable sexaquark $uuddss$ introduced elsewhere and detailed in~\cite{fS22}.   The natural $E_{\rm CM}$ range for \ggbar\ production in these cases is $\approx 2-4.5$ GeV.  

There are currently no other evident candidates to reconcile both the lattice HVP and muon g-2 values with the R-ratio, because to contribute to the lattice HVP, the state must be made of quarks and gluons.  As-yet-unknown colored particles such as color sextet quarks would not do the job because they are not present in the lattice calculation, moreover the contribution to the HVP goes roughly as $R(s)/s^2$ so it is difficult for high mass particles to have sufficient impact.      
The proposed resolution of the discrepancy between muon g-2, lattice QCD, and direct experimental measurement of the R-ratio, will impact electroweak precision sector fits and likely make tensions among those observables more significant~\cite{pms_g-2_08,crivellin+g-2EW20,passera+g-2_20}.


\noindent{\bf Acknowledgements:}\\
I have benefitted from valuable input from Alex Bondar, Marek Karliner, Yury Kolomensky, Arkady Vainshtein, Zihui Wang, Xingchen Xu and Changzheng Yuan.  This research was supported by the National Science Foundation Grant NSF-PHY-2013199 and by the Simons Foundation. 

%


\end{document}